\RequirePackage{fix-cm}
\documentclass[twocolumn,epjc3]{svjour3}          
\smartqed  
\usepackage{amsmath}
\usepackage{graphicx}
\usepackage[utf8]{inputenc}
\begin{document}

\title{Analysis of DIS structure functions of the nucleon within truncated
Mellin moments approach\thanksref[$\star$]{t1}
}
\thankstext[$\star$]{t1}{This work is supported by the Bogo\-liubov -
Infeld Program, Grant No. 01-3-1113-2014/2018.}


\author{D. Kotlorz\thanksref{a,1,2}
        \and
        A. Kotlorz\thanksref{1}
}

\thankstext[$\mu$]{a}{e-mail: dorota@theor.jinr.ru}

\institute{Opole University of Technology,
           Institute of Mathematics and Physics,
           Pr\'oszkowska 76, 45-758 Opole, Poland \label{1}
           \and
           Bogoliubov Laboratory of
           Theoretical Physics, Joint Institute for Nuclear Research,
           Dubna 141980, Russia \label{2}
}

\date{Received: date / Accepted: date}

\maketitle

\begin{abstract}
We present generalized evolution equations and factorization in terms of
the truncated Mellin moments (TMM) of the parton distributions and structure
functions. We illustrate the $x$ and $Q^2$ dependence of TMM in the polarized
case. Using the TMM approach we compare the integrals of $g_1$ with HERMES
and COMPASS data from the limited $x$-ranges.
\end{abstract}

\section{Introduction}
\label{intro}
Our knowledge of the matter structure and fundamental particle interactions
in high energy regimes is mostly provided by deep inelastic scattering (DIS)
of leptons on hadrons and hadron-hadron collisions.
According to the factorization theorem (for a review see, for instance,
\cite{Collins:1989gx}),
the cross sections for DIS and hadron - hadron collisions can be
represented as convolution of short-distance perturbative and long-distance
nonperturbative parts.
The perturbative part describing partonic cross sections at a sufficiently
high scale of the momentum transfer $Q$ is calculable within the
perturbative QCD. In turn, the non-perturbative part contains
universal process independent parton distribution functions (PDFs) and
fragmentation functions (FFs), which can be obtained from experimental data.
The evolution of PDFs and FFs with the interaction scale $Q^2$ is again
described with the use of the perturbative QCD methods.   
Usually, one uses the standard DGLAP approach
\cite{Gribov:1972ri}, \cite{Gribov:1972rt}, \cite{Dokshitzer:1977sg},
\cite{Altarelli:1977zs}
to calculate parton densities at a given scale $Q^2$ when these densities
are assumed for a certain input scale $Q_0^2$.
Traditionally, in the QCD description of the nucleon structure, the central
role is played by the quark and gluon distribution functions and their
evolution equations. Then, Mellin moments of the parton distributions and
structure functions (SFs), which are essential in testing sum rules,
are obtained as integrals of the distribution or structure functions over
the Bjorken-$x$ variable.
An alternative approach, in which one can study directly the evolution of the
truncated moments of the parton distributions was proposed in
\cite{Forte:1998nw}, \cite{Forte:2000wh}, \cite{Piccione:2001vf},
\cite{Forte:2002us}. 
Later on, we elaborated the exact evolution equations
for the truncated moments of the parton densities and structure functions
\cite{Kotlorz:2006dj}, \cite{Kotlorz:2011pk}, \cite{Kotlorz:2014kfa}.
We found that the $n$th truncated Mellin moment obeys the
DGLAP evolution but with the transformed kernel $P(x)'=P(x)x^n$. Also, the
coefficient functions for the truncated moments of the structure functions
have a simple rescaled form $C(x)'=x^nC(x)$. In fact, the TMM approach is
a generalization of the well-known DGLAP evolution for PDFs, where one obtains
the answer for the generalized TMM which can be one of the many possible
constructions, e.g.: PDFs $f(x,Q^2)$, SFs $F(x,Q^2$ themselves, their truncated
or untruncated $n$th moments, multi-integrations or multi-differentiations
of $f$ \cite{Kotlorz:2014fia}.
The major advantage of the TMM approach is a possibility to adapt theoretical
analysis of the nucleon structure functions to the experimentally accessible
region as the measurements do not extend to a very large and a very small 
Bjorken-$x$ variable. Furthermore, solving the evolution equations for
truncated moments, one does not need to assume exact forms of the input
parametrizations of the parton densities, which like, e.g., polarized gluon
distributions, are weakly known. The extraction of the truncated moments of PDFs
or SFs from the data carries smaller uncertainties than the extraction of PDFs
or SFs themselves. The TMM of the original function $f$ for $n\geq 1$ are less
singular in $x$ than $f$ itself and all of the DGLAP evolution and
convolution Wilson kernels, which simply rescale to $x^n P(x)$, $x^n C(x)$,
are also less singular than $P$ and $C$, respectively.
Hence, numerical analysis based directly on the evolution of truncated
moments is faster, more stable and accurate in comparison with the traditional
approach based on PDFs.
These advantages make the TMM approach a promising tool in QCD studies,
providing direct methods to test different unpolarized and polarized sum
rules in each order of perturbation expansion. This is crucial for instance,
in finding out how the nucleon spin is distributed among its constituents:
quarks and gluons.    
A number of important problems in particle physics, e.g., solving of the
mentioned above `nucleon spin puzzle', quark - hadron duality or higher twist
contributions to the structure functions refers directly to moments. 
These issues initiate a large number of experimental and theoretical studies
as well. The TMM approach can be very helpful in these projects.\\

The aim of this paper is to acquaint the Reader with the TMM approach and
encourage Her/Him to take it into account in Her/His own studies.
The content of this paper is as follows. In the next section, we present
the main results of the TMM approach for PDFs and SFs. In Sec.~3, we
illustrate the $Q^2$ evolution of the polarized PDFs and SFs in terms of the
TMM. We also compare the predictions for contributions to the first moment
of the structure function $g_1$ with HERMES and COMPASS data from the limited
$x$-ranges.

\section{Generalized evolution for TMM of the parton distributions and
structure functions}
\label{sec:2}
As it has already been mentioned in the Introduction, the truncated
moments of the original function $f$ in their general form may assume many
different constructions, useful in analysis of the nucleon structure
functions. Each of these TMM obeys the DGLAP evolution with a very simply
rescaled kernel $P_{ij}(x)$. The Reader can find the summarized results in
the Appendix A and more details on this subject in
\cite{Kotlorz:2014fia}, \cite{Strozik-Kotlorz:2015gka},
\cite{Strozik-Kotlorz:2015iqr}.
We also proposed there the generalized Bjorken sum rule as an example of
application of the TMM. Here, in this paper we shall focus on the original
version of the evolution equations for the TMM \cite{Kotlorz:2006dj}.
We shall present the suitable evolution equations and relations for the TMM
of the parton distribution functions in Sec.~\ref{sec:2.1} and structure
functions in Sec.~\ref{sec:2.2}.

\subsection{Evolution of the PDFs and their TMM}
\label{sec:2.1}
In the TMM approach to the DGLAP evolution the main role is played by the
truncated integrals of the original functions $f(x)$,
\begin{equation}\label{eq.1}
f^n(x)\equiv\int\limits_{x}^1 dz\, z^{n-1}\, f(z),
\end{equation}
where $f(x)$ can be any unpolarized $q$ or polarized $\Delta q$ parton
distribution function and $f^n(x)$ defines its $n$th moment truncated at $x$.

Throughout this paper, we use the following notation:
$q(x,Q^2)$, $\bar{q}(x,Q^2)$, $G(x,Q^2)$, $\Delta q(x,Q^2)$,
$\Delta\bar{q}(x,Q^2)$, $\Delta G(x,Q^2)$ denote PDFs, while
$q^n(x,Q^2)$, $\bar{q}^n(x,Q^2)$, $G^n(x,Q^2)$, $\Delta q^n(x,Q^2)$,
$\Delta\bar{q}^n(x,Q^2)$, $\Delta G^n(x,Q^2)$ are their TMM,
defined as in Eq.~(\ref{eq.1}), respectively.

The well-known DGLAP evolution equations for the nonsinglet distributions
$q_{NS}$ take the form
\begin{equation}\label{eq.2}
\frac{\partial}{\partial\ln Q^2} q_{NS}(x,Q^2)=\frac{\alpha_s(Q^2)}{2\pi}\;
(P_{qq}\ast q_{NS})(x,Q^2)
\end{equation}
and for the singlet $q_S$ and gluon distributions $G$ they are the matrix
equation,
\begin{eqnarray}\label{eq.3}
&&\frac{\partial}{\partial\ln Q^2}
\begin{pmatrix} 
q_{S}(x,Q^2) \\
G(x,Q^2)
\end{pmatrix}=\nonumber \\
&&\frac{\alpha_s(Q^2)}{2\pi}\left(
\begin{pmatrix}
P_{qq} & P_{qG}\\         
P_{Gq} & P_{GG}               
\end{pmatrix}
\ast
\begin{pmatrix} 
q_{S}\\
G   
\end{pmatrix}\right)(x,Q^2).
\end{eqnarray}
In the above equations, $\ast$ denotes the Mellin convolution,
\begin{equation}\label{eq.4}
(P\ast q)(x,Q^2)\equiv 
\int\limits_{x}^1 \frac{dz}{z}\; P\left(\frac{x}{z}\right)q(z,Q^2).
\end{equation}
Each splitting function $P_{ij}$ is calculable as a power
series in the strong coupling constant $\alpha_s$,
\begin{equation}\label{eq.5}
P_{ij}(x,\alpha_s) = \left [\, P_{ij}^{(0)}(x) + \frac{\alpha_s(Q^2)}{2\pi}
P_{ij}^{(1)}(x) + \cdots \, \right].
\end{equation}
For the polarized parton densities $\Delta q_i(x,Q^2)$,
$\Delta G(x,Q^2)$ the evolution equations have the same form,
Eqs.~(\ref{eq.2}), (\ref{eq.3}), but with the polarized splitting
functions $\Delta P_{ij}(x,\alpha_s)$, respectively.

Taking into account the properties of the Mellin convolution and the basic
physical condition that parton densities disappear for $x>1$, we found in
\cite{Kotlorz:2006dj} that the truncated moments of the parton distributions
defined in Eq.~(\ref{eq.1}) also obey the DGLAP evolution equations with
slightly modified evolution kernels, namely
\begin{equation}\label{eq.6}
\frac{\partial}{\partial\ln Q^2} q^n_{NS}(x,Q^2)=\frac{\alpha_s(Q^2)}{2\pi}\;
(P'_{qq}\ast q^n_{NS})(x,Q^2),
\end{equation}
\begin{eqnarray}\label{eq.7}
&&\frac{\partial}{\partial\ln Q^2}
\begin{pmatrix} 
q^n_{S}(x,Q^2) \\
G^n(x,Q^2)
\end{pmatrix}
=\nonumber \\
&&\frac{\alpha_s(Q^2)}{2\pi}\left(
\begin{pmatrix}
P'_{qq} & P'_{qG}\\         
P'_{Gq} & P'_{GG}               
\end{pmatrix}
\ast
\begin{pmatrix} 
q^n_{S}\\
G^n   
\end{pmatrix}\right)(x,Q^2),
\end{eqnarray}
\nopagebreak
where
\begin{equation}\label{eq.8}
P'_{ij}(x,\alpha_s) = x^n P_{ij}(x,\alpha_s).
\end{equation}   
Similar equations hold in the polarized case:
\begin{equation}\label{eq.9}
\frac{\partial}{\partial\ln Q^2} \Delta q^n_{NS}(x,Q^2)=
\frac{\alpha_s(Q^2)}{2\pi}\;(\Delta P'_{qq}\ast \Delta q^n_{NS})(x,Q^2),
\end{equation}
\begin{eqnarray}\label{eq.10}
&&\frac{\partial}{\partial\ln Q^2}
\begin{pmatrix} 
\Delta q^n_{S}(x,Q^2) \\
\Delta G^n(x,Q^2)
\end{pmatrix}=\nonumber \\
&&\frac{\alpha_s(Q^2)}{2\pi}\left(
\begin{pmatrix}
\Delta P'_{qq} & \Delta P'_{qG}\\         
\Delta P'_{Gq} & \Delta P'_{GG}               
\end{pmatrix}
\ast
\begin{pmatrix} 
\Delta q^n_{S}\\
\Delta G^n   
\end{pmatrix}\right)(x,Q^2),
\end{eqnarray}
where again as in Eq.~(\ref{eq.8})
\begin{equation}\label{eq.11}
\Delta P'_{ij}(x,\alpha_s) = x^n \Delta P_{ij}(x,\alpha_s).
\end{equation}
Since existing measurements cover only a restricted range in $x$,
$x_{min}\leq x\leq x_{max}$, it is useful to consider the double truncated
Mellin moments of PDFs which are defined by:
\begin{equation}\label{eq.12}
f^{n}(x_{1},x_{2}) = 
\int\limits_{x_{1}}^{x_{2}} dx\; x^{n-1}\,f(x).
\end{equation} 
It is straightforward to show that the double truncated moments,
Eq.~(\ref{eq.12}), being a subtraction of two single truncated ones,
\begin{equation}\label{eq.13}
f^{n}(x_{1},x_{2}) = f^{n}(x_{1}) - f^{n}(x_{2}),  
\end{equation}
also fulfill the analogical DGLAP equations
\cite{Psaker:2008ju}, \cite{Kotlorz:2011pk}:
\begin{eqnarray}\label{eq.14}
&&\frac{\partial}{\partial t} q^{n}_{NS}(x_{1},x_{2},Q^2)=\nonumber \\
&&\frac{\alpha_s(Q^2)}{2\pi}\;
\int\limits_{x_{1}}^{1}\frac{dz}{z}\; P'(z)\;
q^{n}_{NS}\left( \frac{x_{1}}{z}, \frac{x_{2}}{z},Q^2 \right).
\end{eqnarray}
Notice that the evolution equations for the double truncated moments
Eq.~(\ref{eq.14}) are in fact a generalization of those for
the single truncated and untruncated ones. Setting $x_{2}=1$
one obtains Eq.~(\ref{eq.6}), while setting $x_{1}=0$ and $x_{2}=1$
one obtains the well-known renormalization group equations for the untruncated
moments:
\begin{equation}\label{eq.15}
\frac{\partial}{\partial\ln Q^2} q^{n}_{NS}(Q^2)=
\frac{\alpha_s(Q^2)}{2\pi}\,\gamma^{n}_{qq}(Q^2)q^{n}_{NS}(Q^2).
\end{equation}

\subsection{Evolution of SFs and their TMM}
\label{sec:2.2}
Similarly to the evolution of the TMM of PDFs, where the splitting functions
have a simply modified form, Eq.~(\ref{eq.8}), the coefficient functions of
the $n$th truncated moments for structure functions are changed in the same
manner \cite{Kotlorz:2014kfa}. Namely, if $F$ denotes SF
\begin{equation}\label{eq.16}
F(x) = (C*f)(x),
\end{equation}
then the TMM of $F$,
\begin{equation}\label{eq.17}
F^n(x)\equiv\int\limits_x^1 z^{n-1}F(z)\, dz
\end{equation}
takes the form
\begin{equation}\label{eq.18}
F^n(x) = (C'*f^n)(x),
\end{equation}
where $f^n$ is the TMM of PDF $f$ defined in Eq.~(\ref{eq.1}) and $C'$ is
the new coefficient function
\begin{equation}\label{eq.19}
C'(x) = x^n C(x).
\end{equation}
Let us demonstrate this for a case of the polarized structure function $g_1$.
 
In the NLO approximation within the ${\bar MS}$ scheme $g_1(x,Q^2)$ is given by
\begin{eqnarray}\label{eq.20}
g_1(x,Q^2)&=&\frac{1}{2}\sum_q e_q^2\Big[\Delta q(x,Q^2) +
\Delta\bar{q} (x,Q^2)\nonumber \\
&+&\frac{\alpha_s(Q^2)}{2\pi}\Big( (\Delta C_q *
(\Delta q + \Delta\bar{q}))(x,Q^2)\nonumber \\
&+& (2\Delta C_G * \Delta G)(x,Q^2)\Big)\Big],
\end{eqnarray}
where $\Delta C_i$ denotes the spin dependent coefficient functions and
$\Delta\bar{q}$ is the antiquark polarized PDF.
According to Eqs.~(\ref{eq.16})--(\ref{eq.19}), the $n$th TMM of $g_1$,
\begin{equation}\label{eq.21}
g_1^n(x,Q^2) = \int\limits_x^1 z^{n-1}g_1(z,Q^2)\, dz
\end{equation}
obtains the following form:
\begin{eqnarray}\label{eq.22}
g_1^n(x,Q^2)&=&\frac{1}{2}\sum_q e_q^2\Big[\Delta q^n(x,Q^2) +
\Delta\bar{q}^n (x,Q^2)\nonumber \\
&+&\frac{\alpha_s(Q^2)}{2\pi}\Big( (\Delta C'_q *
(\Delta q^n + \Delta\bar{q}^n))(x,Q^2)\nonumber \\
&+& (2\Delta C'_G * \Delta G^n)(x,Q^2)\Big)\Big]
\end{eqnarray}
where
\begin{equation}\label{eq.23}
\Delta C'_{q,G}(x) = x^n \Delta C_{q,G}(x).
\end{equation}

Finally, we also would like to mention some results for the function $g_2$,
implied by the TMM approach.

In \cite{Kotlorz:2011pk}, we derived the Wandzura-Wilczek (WW) relation
\cite{Wandzura:1977qf} for the TMM, and found partial contributions to
the Burkhardt-Cottingham (BC) \cite{Burkhardt:1970ti} sum rule.
Namely, the WW relation in terms of the TMM reads
\begin{equation}\label{eq.24}
g_2^n(x,Q^2) = \frac{1-n}{n}\: g_1^n(x,Q^2) - \frac{x^n}{n}\:g_1^0(x,Q^2), 
\end{equation}
where
\begin{equation}\label{eq.25}
g_{1,2}^n(x,Q^2)=\int\limits_{x}^1 dz\, z^{n-1}\, g_{1,2}(z,Q^2).
\end{equation}
In a case of the first moment ($n=1$), from Eq.~(\ref{eq.24}) one gets
\begin{eqnarray}\label{eq.26}
&&\int\limits_{x_1}^{x_2} dx\, g_2^{WW}(x,Q^2)=\nonumber \\
&&-x_1\int\limits_{x_1}^{x_2}\frac{dz}{z}\, g_1(z,Q^2)+
(x_2-x_1)\int\limits_{x_2}^1\frac{dz}{z}\,g_1(z,Q^2).
\end{eqnarray}
The above equations can be used to testing the BC sum rule and other TMM of
$g_2$.

The advantage of the TMM approach is the possibility to have a convenient
procedure that combines direct evolution of important physical quantities
with factorization in a smaller number of steps than the widely-known approach
based on PDFs.
Furthermore, since the suitable functions in the TMM approach are mostly
more regular (less singular for very small $x$) than in the case of PDFs,
the numerical procedures used in the TMM approach are more stable than
those for the standard PDFs approach.

\section{TMM results for the spin PDFs and SFs}
\label{sec:3}
One of the most important goals in the recent studies of QCD is
understanding of the nucleon spin structure and determination of the
individual partonic contributions to the helicity of the nucleon,
\begin{equation}\label{eq.27}
\Delta q(Q^2) = \int\limits_0^1 dx\, \Delta q(x,Q^2).
\end{equation}
Measurements of the spin structure functions $g_1$ and $g_2$, which
parametrize the cross section of polarized inclusive DIS, are always
performed in the restricted $x$-range. This limitation provides results,
which are the partial (truncated) moments of the parton helicity distributions,
\begin{equation}\label{eq.28}
\Delta q(x_1,x_2,Q^2) = \int\limits_{x_1}^{x_2} dx\, \Delta q(x,Q^2),
\end{equation}
instead of the full moments, Eq.~(\ref{eq.27}). Thus, the TMM, and
especially the first TMM of the polarized PDFs and SFs, are quantities of
large importance. The approach, presented here, allows one a direct
study of these quantities.

Here we present the numerical results for evolution of the TMM.
In Figs. 1--10, we illustrate the $x$ and $Q^2$ dependence of
the TMM of the polarized PDFs and SFs in LO and NLO.
We solve the evolution Eqs.~(\ref{eq.9})--(\ref{eq.11}) in the $x$-space and
also use the factorization formula for SF $g_1$,
Eqs.~(\ref{eq.22})--(\ref{eq.23}) (for more details, see Appendix B).
Finally, in  Table~\ref{table 1}, we present a comparison with HERMES
\cite{Airapetian:2006vy} and COMPASS \cite{Adolph:2015saz} data on the first
TMM of the spin SF $g_1$. We show results for the partial contributions to
the integrals of $g_1$,
\begin{equation}\label{eq.29}
\Gamma_1(x_1,x_2,Q^2)=\int\limits_{x_1}^{x_2} dx\, g_1(x,Q^2),
\end{equation}
for the proton, neutron, deuteron, nucleon and the nonsinglet part.
The truncated contribution to the nonsinglet SF,
\begin{equation}\label{eq.30}
\int\limits_{x_1}^{x_2} dx\, g_{1,NS}(x,Q^2)=
\int\limits_{x_1}^{x_2} dx\, (g_{1,p} - g_{1,n})(x,Q^2)
\end{equation}
is crucial in determination of the Bjorken Sum Rule (BSR)
\cite{Bjorken:1966jh}, \cite{Bjorken:1969mm}.
\begin{figure}[!h] 
\includegraphics[width=0.50\textwidth]{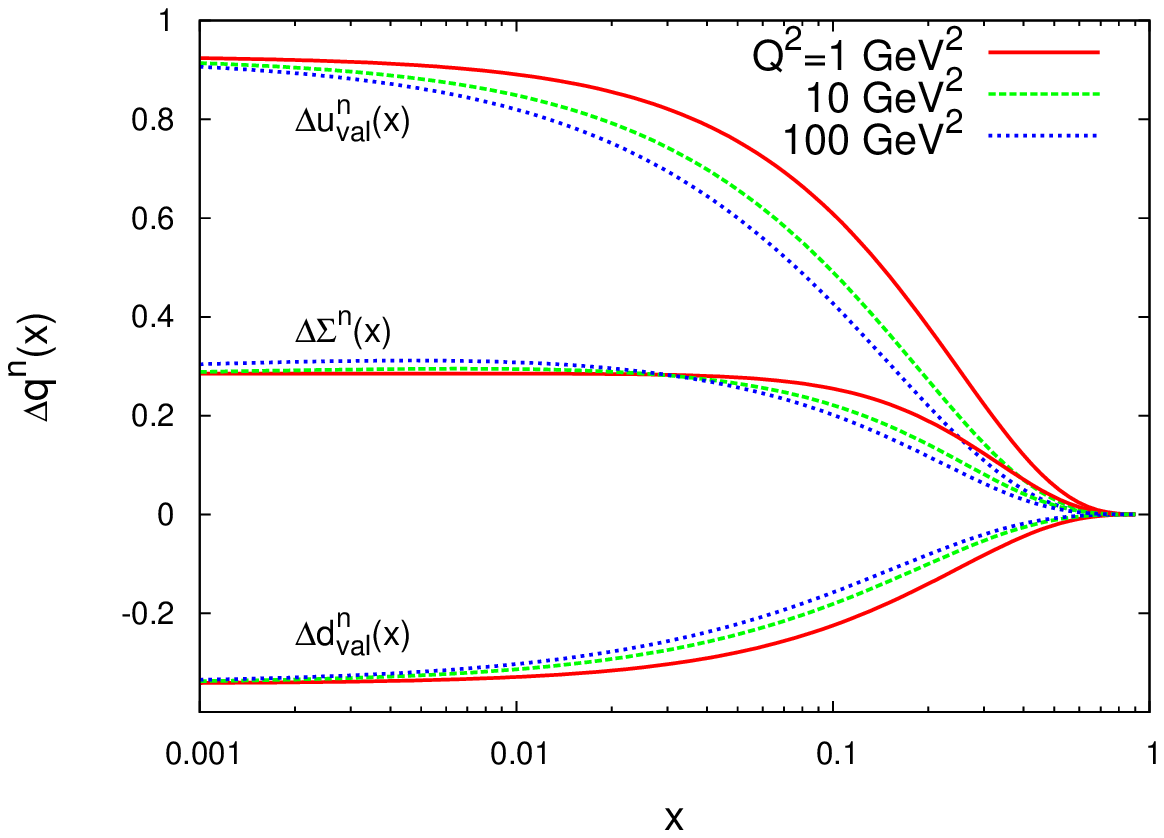}
\includegraphics[width=0.50\textwidth]{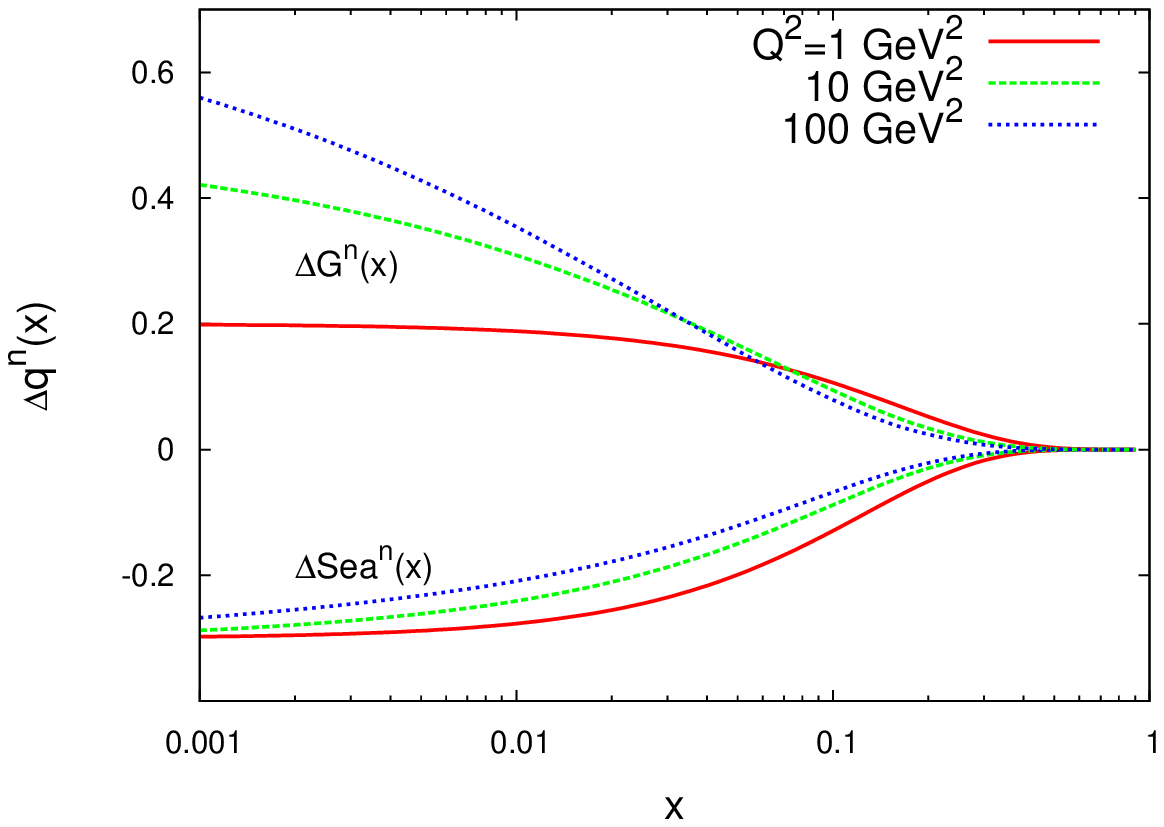}
\caption{The first TMM ($n=1$) of the polarized PDFs,
$\int\limits_x^1 dz\, \Delta q(z,Q^2)$, as a~function of the
low-$x$ limit of integration, at different $Q^2$, in NLO.}
\label{fig1}
\end{figure}
\begin{figure}[!h] 
\includegraphics[width=0.50\textwidth]{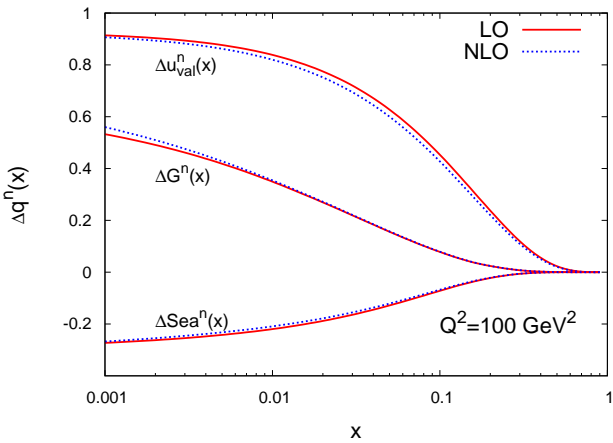}
\caption{A comparison of LO and NLO evolution of the first TMM of
$\Delta u_{val}$, $\Delta Sea$, $\Delta G$, vs the
low-$x$ limit of integration, at $Q^2=100\;\rm{GeV}^2$.}
\label{fig2}
\end{figure}
\begin{figure}[!h] 
\includegraphics[width=0.50\textwidth]{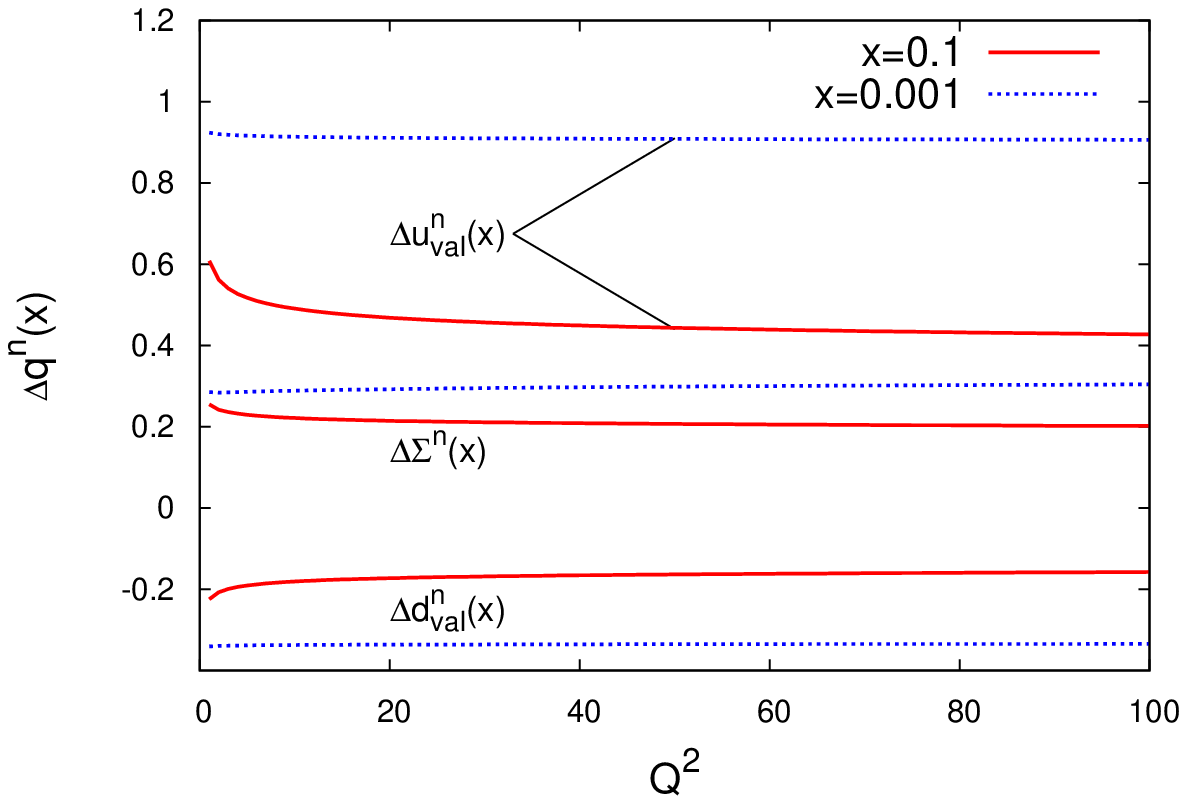}
\includegraphics[width=0.50\textwidth]{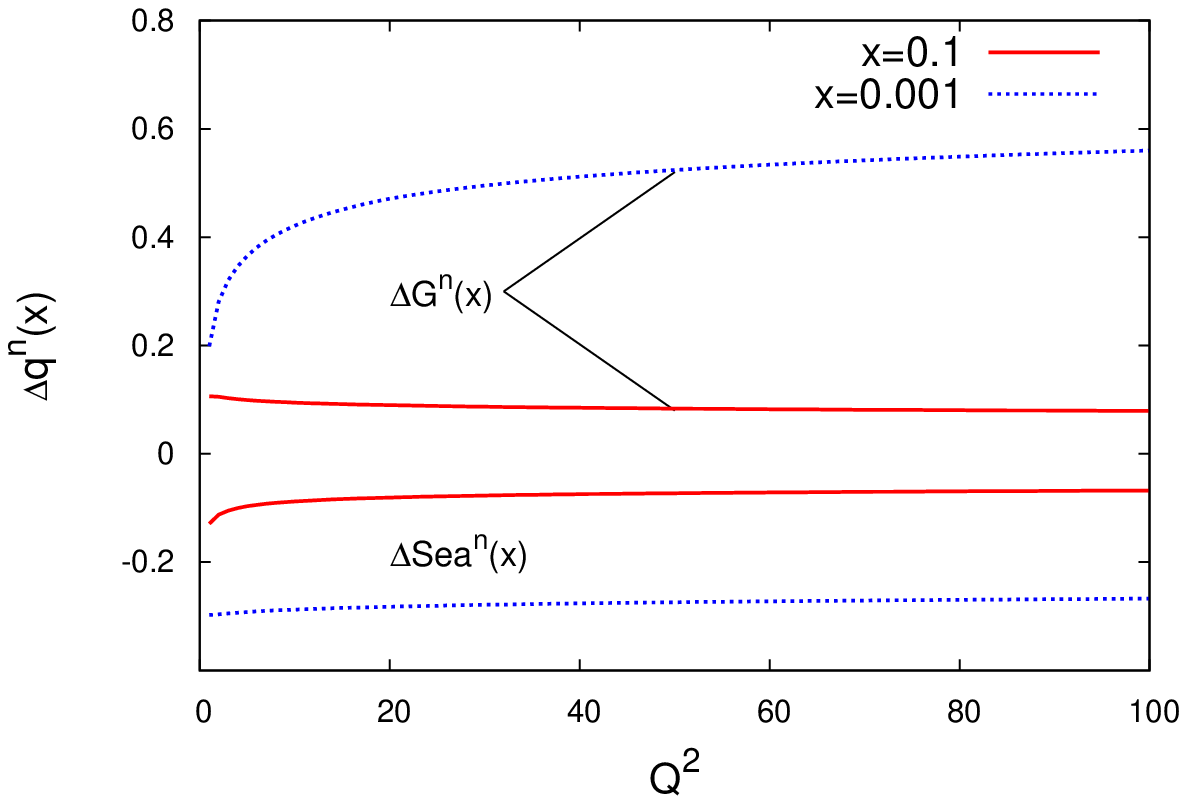}
\caption{The first TMM of the polarized PDFs, as a~function of $Q^2$,
for two low-$x$ limits of integration, in NLO.}
\label{fig3}
\end{figure}
\begin{figure}[!h] 
\includegraphics[width=0.50\textwidth]{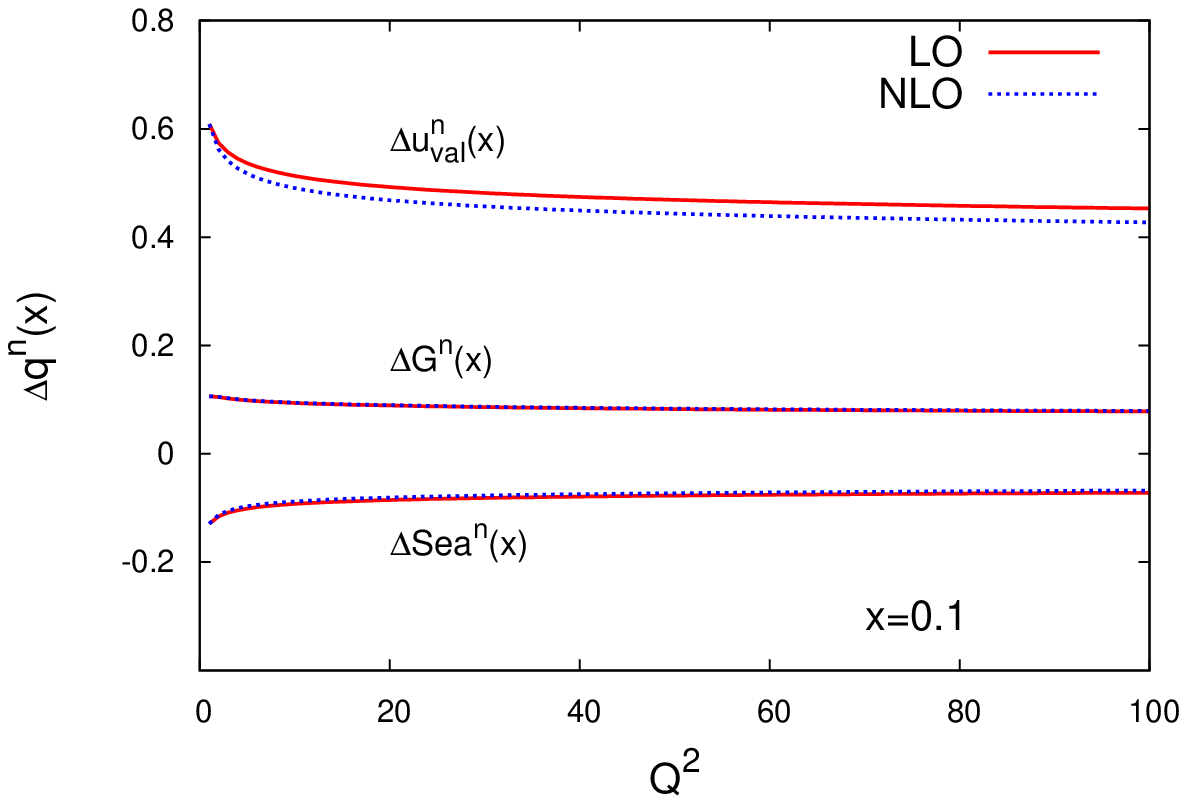}
\includegraphics[width=0.50\textwidth]{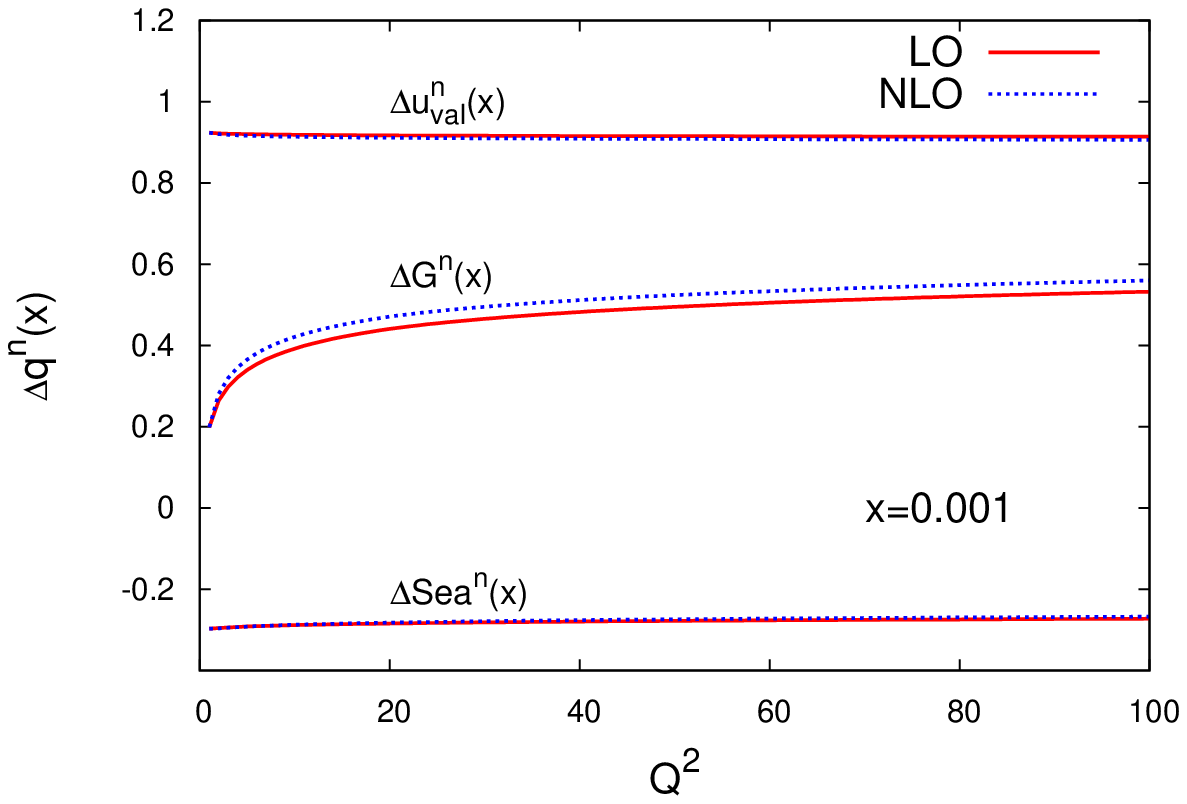}
\caption{A comparison of LO and NLO evolution of the first TMM of
$\Delta u_{val}$, $\Delta Sea$ and $\Delta G$, vs $Q^2$, at two
low-$x$ limits of integration.}
\label{fig4}
\end{figure}
\begin{figure}[!h] 
\includegraphics[width=0.50\textwidth]{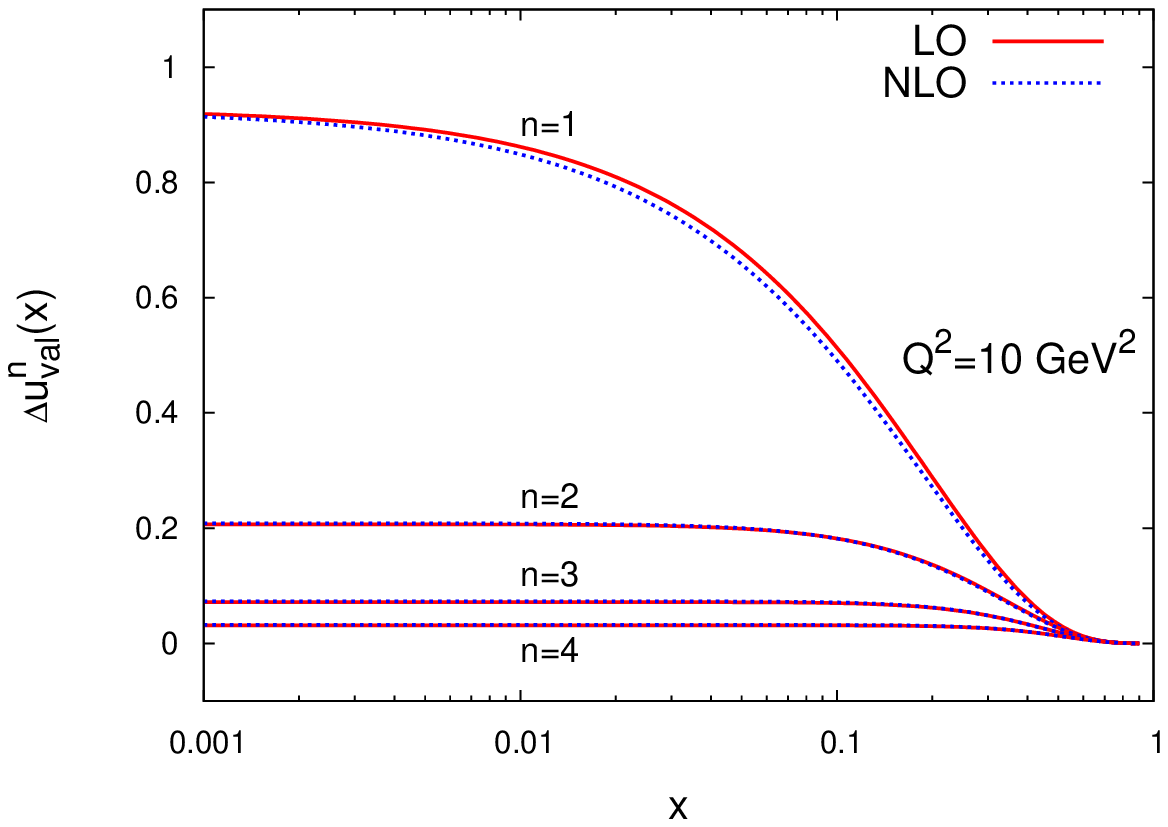}
\includegraphics[width=0.50\textwidth]{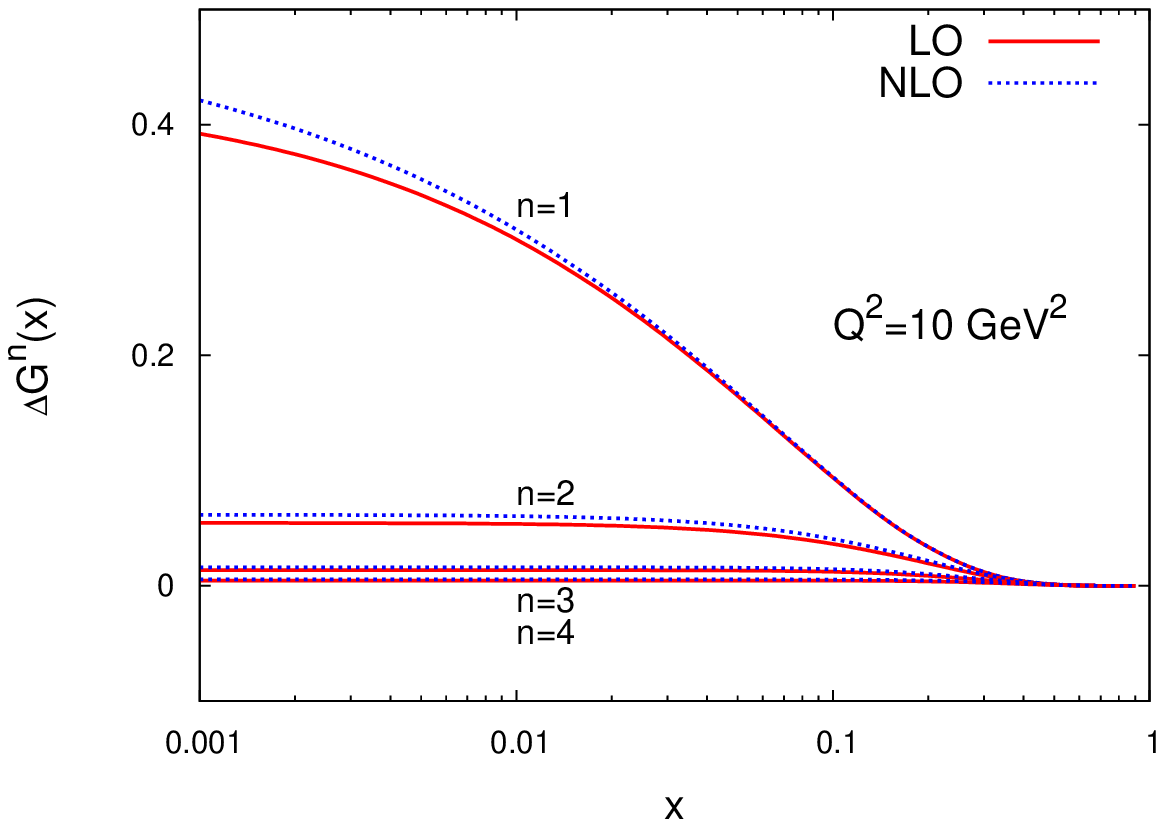}
\caption{A comparison of LO and NLO evolution of four TMMs ($n=1,\,2,\,3,\,4$)
of $\Delta u_{val}$ and $\Delta G$,
vs low-$x$ limit of integration, at $Q^2=10\;\rm{GeV}^2$.}
\label{fig5}
\end{figure}
\begin{figure}[!h] 
\includegraphics[width=0.50\textwidth]{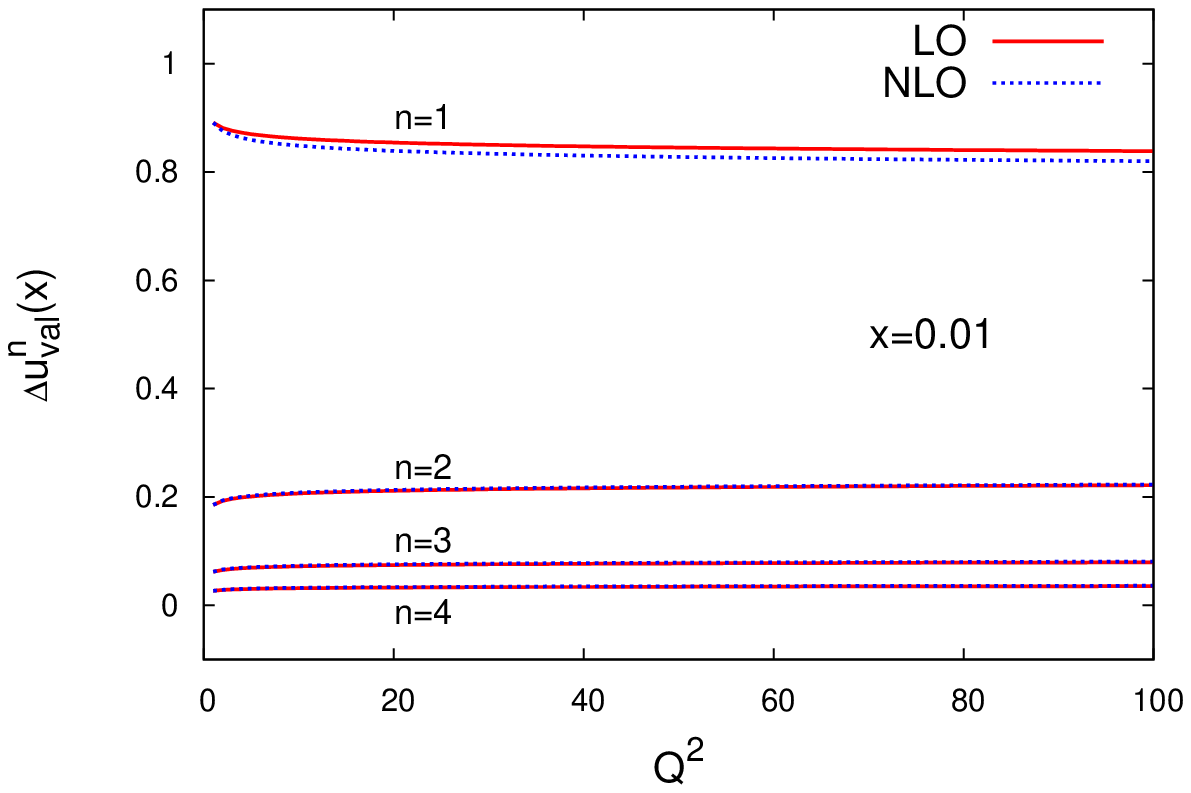}
\includegraphics[width=0.50\textwidth]{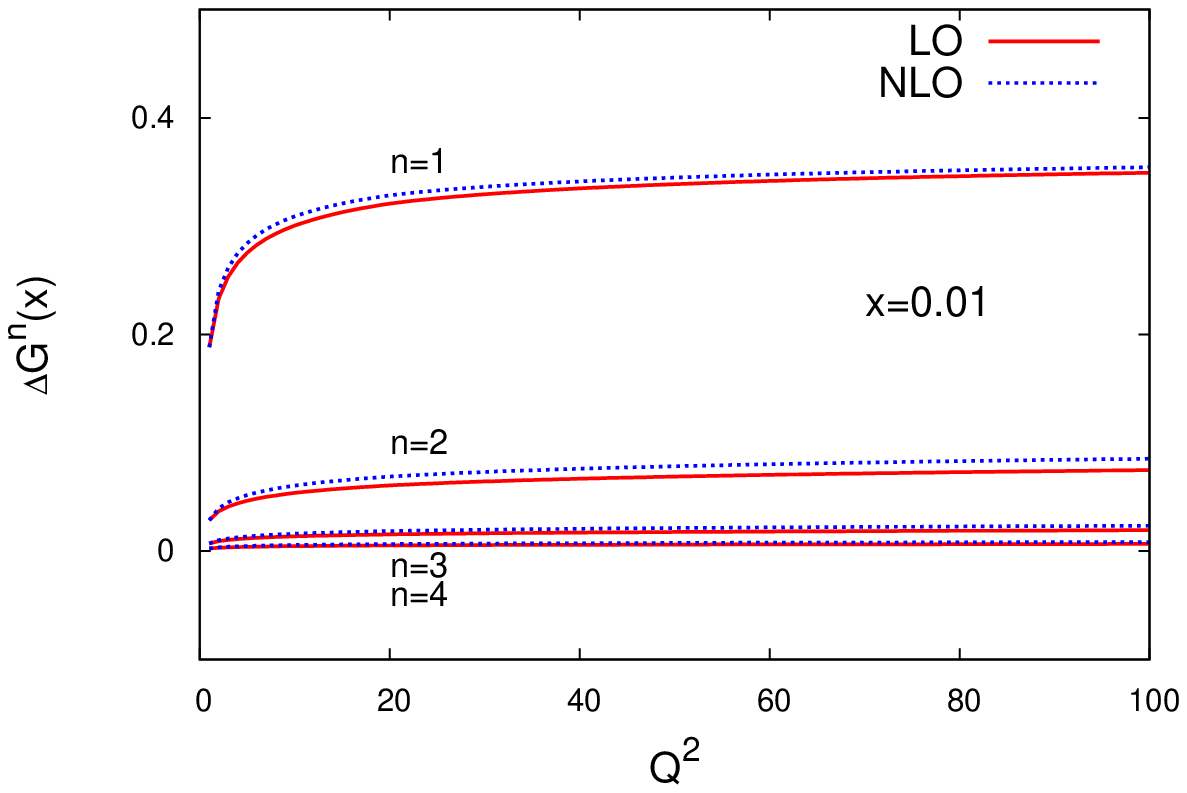}
\caption{A comparison of LO and NLO evolution of four TMMs ($n=1,\,2,\,3,\,4$)
of $\Delta u_{val}$ and $\Delta G$,
vs $Q^2$, at low-$x$ limit of integration 0.01.}
\label{fig6}
\end{figure}
\begin{figure}[!h] 
\includegraphics[width=0.50\textwidth]{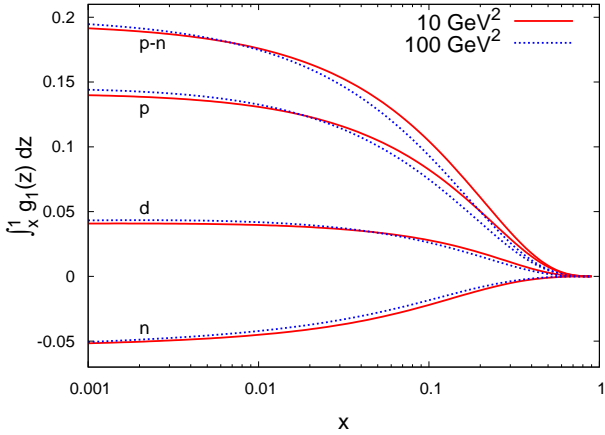}
\caption{The first TMM of the polarized SF $g_1$, for the proton ($p$),
deuteron ($d$), neutron ($n$) and the nonsinglet part ($NS=p-n$),
$\int\limits_x^1 dz\, g_1(z,Q^2)$, as a~function of the
low-$x$ limit of integration, for two values of $Q^2$, in NLO.}
\label{fig7}
\end{figure}
\begin{figure}[!h] 
\includegraphics[width=0.50\textwidth]{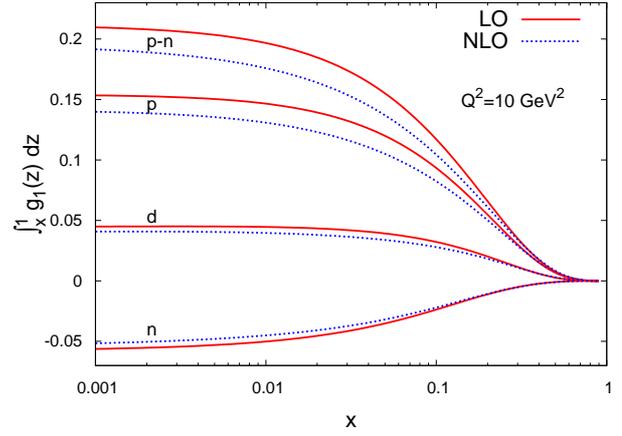}
\caption{A comparison of LO and NLO evolution of the first TMM of
$g_1$, for the proton, deuteron, neutron and the non\-singlet part,
as a~function of the low-$x$ limit of integration, at $Q^2=10\;\rm{GeV}^2$.}
\label{fig8}
\end{figure}
\begin{figure}[!h] 
\includegraphics[width=0.50\textwidth]{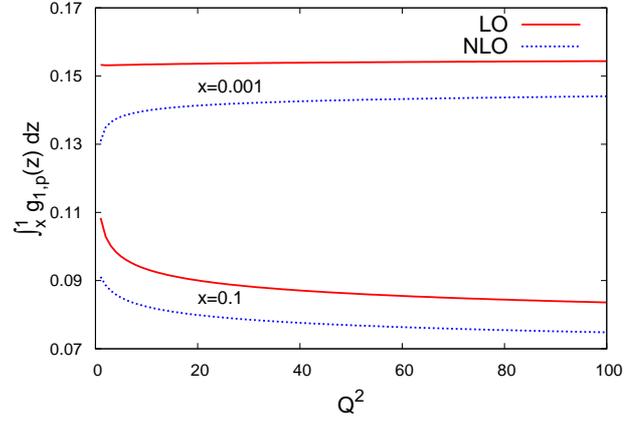}
\includegraphics[width=0.50\textwidth]{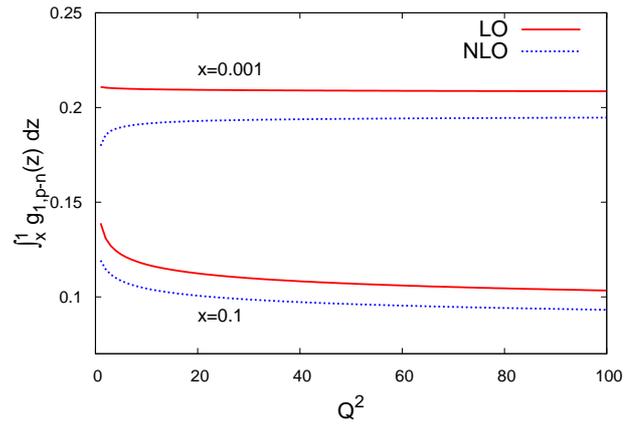}
\caption{A comparison of LO and NLO evolution of the first TMM of
$g_1$, for the proton and the nonsinglet part (p-n),
vs $Q^2$, for two low-$x$ limits of integration.}
\label{fig9}
\end{figure}
\begin{figure*}[!h] 
\centering
\includegraphics[width=0.45\textwidth]{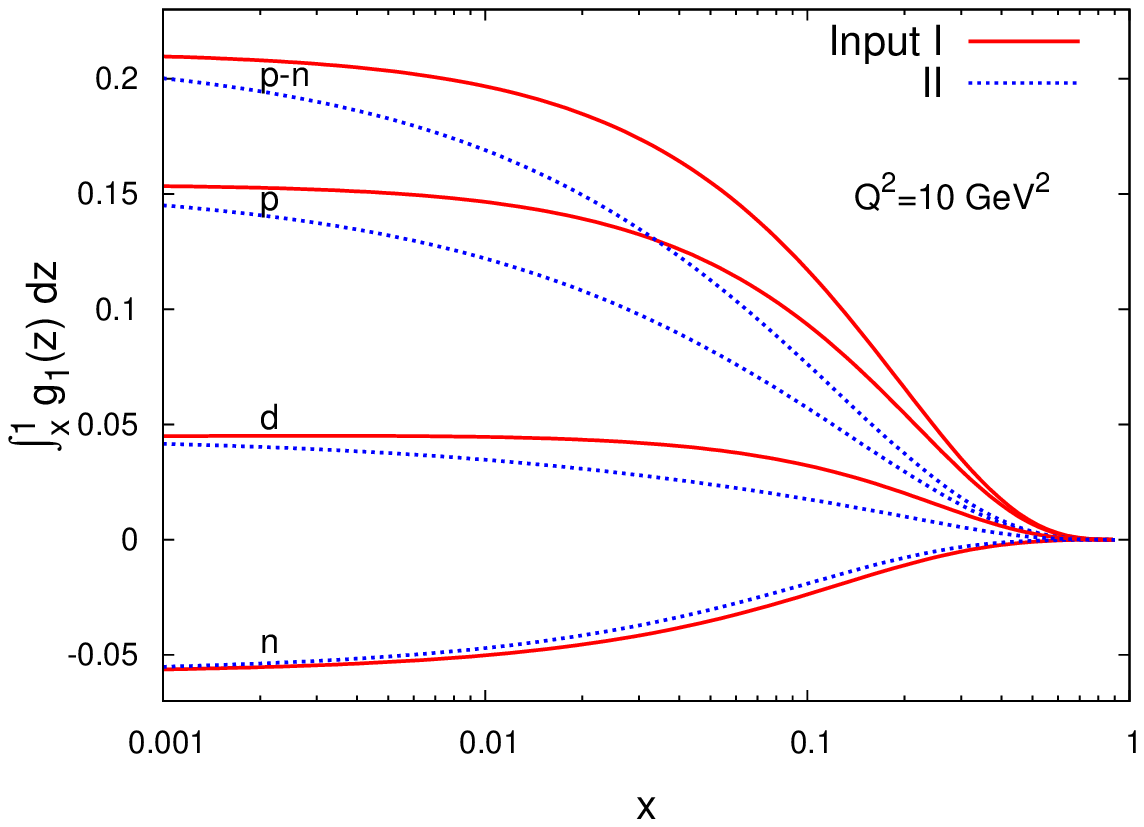}
\includegraphics[width=0.45\textwidth]{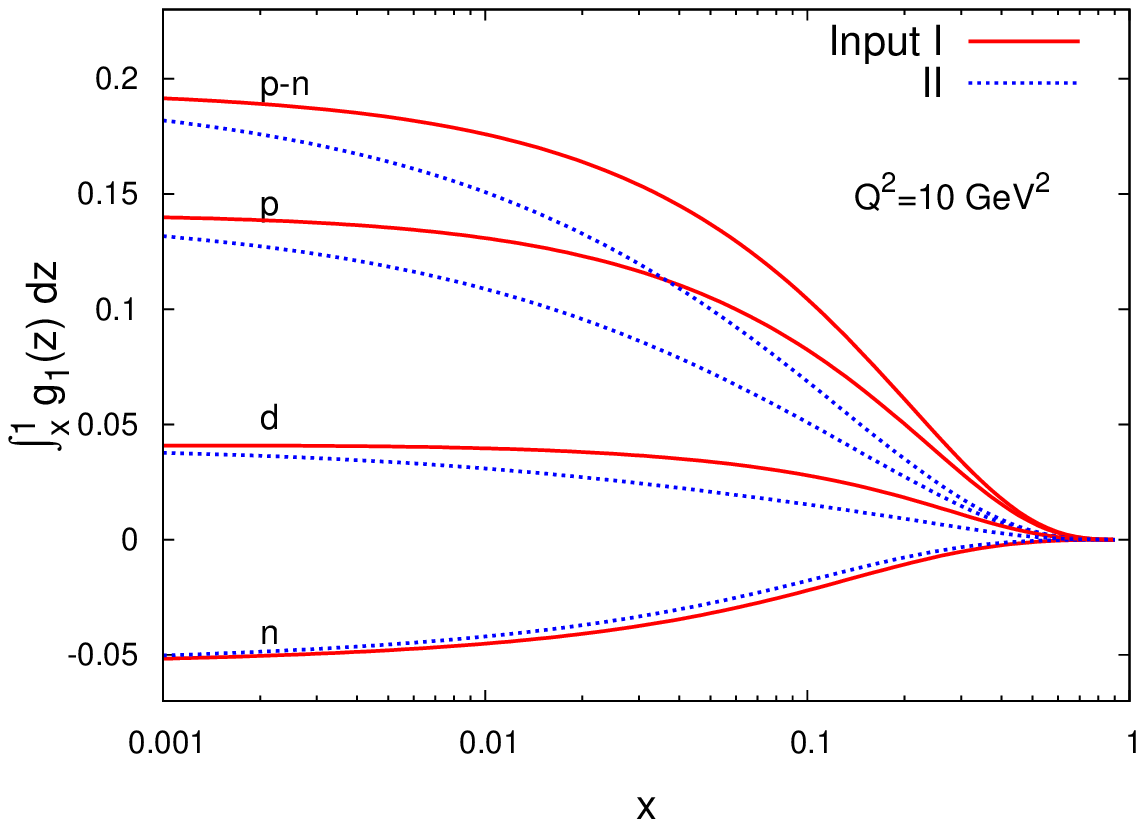}
\caption{A comparison of the impact of the different small-$x$ behaviour of
the polarized nonsinglet PDFs on the LO (left) and NLO (right) evolution of
the first TMM of $g_1$. The details on Input~I and Input~II are given in
Appendix B.}
\label{fig10}
\end{figure*}
\begin{table*}[!h]
\caption{\label{table 1}First TMM of $g_1$ in NLO. Comparison with HERMES
\cite{Airapetian:2006vy} and COMPASS \cite{Adolph:2015saz} data.
$g_{1,N}=\frac{1}{2}(g_{1,p}+g_{1,n})$.}
\begin{center}
\begin{tabular*}{\textwidth}{@{\extracolsep{\fill}}lllll@{}}
\hline
Experiment  & Type & Exp. value & Input I & Input II    \\
\hline
 & & & & \\
HERMES & proton & $0.1211\pm 0.0025\pm 0.0068$ & 0.1220 & 0.09513 \\
$Q^2=5\;\rm{GeV}^2$ & deuteron & $0.0436\pm 0.0012\pm 0.0018$ & 0.03724 & 0.02651 \\
$x$-range: 0.021 -- 0.9 & p-n & $0.1479\pm 0.0055\pm 0.0142$ & 0.1635 & 0.1329 \\
 & & & & \\
COMPASS & proton & $0.134\pm 0.003$ & 0.1334 & 0.1229 \\
$Q^2=3\;\rm{GeV}^2$ & N & $0.047\pm 0.003$ & 0.04186 & 0.03737 \\
$x$-range: 0.0025 -- 0.7 & p-n & $0.170\pm 0.008$ & 0.1832 & 0.1710 \\
 & & & & \\
\hline
\end{tabular*}
\end{center}
\end{table*}

\section{Summary}
Our goal in this paper was to present the TMM approach as a convenient
tool in QCD analysis that combines direct evolution of important physical
quantities with factorization in a smaller number of steps than the standard
approach based on PDFs.
Splitting functions $P'$ and coefficient functions $C'$ for the TMM have
simple forms $P'=x^n P$ and $C'=x^n C$, which enables one to use the standard
methods of solving the DGLAP equations only with tiny modifications.
From the technical point of view, the TMM less suffer from experimental
uncertainties, and also the numerical procedures involved into the TMM approach
are more stable than those for PDFs.
The TMM approach is, on the one hand, a generalization of the DGLAP
evolution and, on the other hand, allows a better fit of the theoretical
methods to the limitations of experimental measurements on the kinematic
variables $x$ and $Q^2$. The perturbative QCD itself explores truncated
evolution in $Q^2>\mu^2$; also the Bjorken variable $x\rightarrow 0$ has no
physical meaning (it means infinite energy). Hence, the use of the methods,
which incorporate these limi\-tations in a natural way is very
advantageous.

\begin{acknowledgements}
Special thanks to Michael
\end{acknowledgements}

\appendix

\section{Generalized evolution DGLAP}\label{appA}
The TMM of the parton densities, Eq.~(\ref{eq.1}), and also the generalized
truncated moments obtained by multiple integrations as well as multiple
differentiations of the original parton distribution satisfy the DGLAP
equations with the simply transformed evolution kernel
\cite{Kotlorz:2006dj}, \cite{Kotlorz:2011pk}, \cite{Kotlorz:2014kfa},
\cite{Kotlorz:2014fia}.
In Table~\ref{table 2}, we summarize the generalized TMM together with the
correspondingly transformed DGLAP evolution kernels.
\begin{table*}[!h]
\caption{\label{table 2}TMM and the corresponding evolution kernels}
\begin{center}
\begin{tabular*}{\textwidth}{@{\extracolsep{\fill}}lll@{}}
\hline
Description  & Generalized form & DGLAP evolution kernel ${\cal P}$    \\
\hline
  &                                               &                  \\
Original PDF
  &   $f(x)$                                      & $P(y)$           \\
  &                                               &                  \\
$n$th TMM of PDF
  &   $\int\limits_{z}^{1} x^{n-1}\,f(x)\,dx$             & $P(y)\cdot y^n$  \\
  &                                               &                  \\
Multiple integration 
&   $\int\limits_z^1 z_{k}^{n_k-1} dz_k \: ...
\int\limits_{z_{2}}^{1} z_{1}^{n_1-1}\;f(z_1)\;dz_1$
& $P(y)\cdot y^{n_1+n_2+...+n_k}$  \\
  &                                               &                  \\
Multiple differentiation 
&$\left( -\frac{d}{dx}\right)^k\left[x^n f(x)\right]$& $P(y)\cdot y^{n-k}$\\
  &                                               &                  \\
Convolution with\\
normalized function $\omega(x)$, &$\omega \ast fx^n\equiv
\int\limits_{z}^{1} x^{n-1}\,\omega \left(z/x\right)\, f(x)\,dx$
& $P(y)\cdot y^{n}$ \\
$\int\limits_{0}^{1}\omega(t)\, dt = 1$\\
\hline
\end{tabular*}
\end{center}
\end{table*}

\section{The direct solving of the evolution equations for TMM}\label{appB}
For a fixed $n$, the truncated moment of $f$ Eq.~(\ref{eq.1}) is, like the
$f$ itself, a function of two variables: $x$ - the lower limit of the
integration and $Q^2$. The similarity of the evolution equations for the TMM,
Eqs.~(\ref{eq.6})--(\ref{eq.8}), to the ordinary DGLAP for PDFs,
Eqs.~(\ref{eq.2})--(\ref{eq.4}), enables one to use the same methods of
solving in both the cases.
In literature, there are two basic methods of solving the DGLAP evolution
equations for the function $f$: in the $x$ space with the help of the
polynomial expansions of $f$, (see e.g. \cite{Kumano:2004dw}), or in the
moment space. The use of the moment space gives the possibility to get
analytical solutions for the moments and then, the function $f$ can be
obtained via the inverse Mellin transform. Solving the evolution equation for
the TMM in the $n$ space, one encounters the objects `moment of moment' and the
problem how to deal with them. In \cite{Kotlorz:2011pk}, we derived for this
aim useful relations between untruncated and truncated Mellin moments.

In many our previous TMM analyses we used the Chebyshev polynomials expansion
which is one of the methods of solving the DGLAP equations in the $x$ space,
reducing the former differentio-integral equations to a system of linear
differential ones \cite{El-gendi:1969}.
In this work, for carrying out the DGLAP evolution for the TMM in the
$x$-space we adapted the Hoppet package \cite{Salam:2008qg}, which we
appropriately changed.
As an example, we solve the polarized case of evolution,
Eqs.~(\ref{eq.9})--(\ref{eq.11}), with input truncated moments at
$Q_0^2=1 \rm{GeV}^2$,
\begin{equation}\label{eq.B.1}
\Delta q^n(x,Q_0^2)\equiv\int\limits_{x}^1 dz\, z^{n-1}\, \Delta q(z,Q_0^2).
\end{equation}
We assume distributions $\Delta q(z,Q_0^2)$ in the form
\begin{equation}\label{eq.B.2}
\Delta q_i(x,Q_0^2) =
\frac{N_i x^{a_i}(1-x)^{b_i}}{\int\limits_0^1 x^{a_i}(1-x)^{b_i}\, dx}\, ,
\end{equation}
where $a_{\bar s}=a_G=0$, $b_{uval}=b_{dval}=3$, $b_{\bar s}=7$, $b_G=5$.
In Input~I, we also assume $a_{uval}=a_{dval}=0$, while in Input~II,
$a_{uval}=a_{dval}=-0.4$, that results from theoretical studies on the
small-$x$ behaviour of the nonsinglet polarized PDFs
\cite{Kwiecinski:1995rm}, \cite{Bartels:1995iu}.
The normalization factors $N_i$ reflect the experimental data on the
proton spin contributions: $N_{uval}+N_{dval}=0.585$,
$N_{uval}-N_{dval}=1.270$, $2N_{\bar s}=-0.10$, $N_G=0.2$.

Instead of the the functional form of input PDFs in order to create the
initial TMM, Eq.~(\ref{eq.B.1}), one can represent directly the TMM on a
$x$-space grid.
\clearpage

\bibliographystyle{spphys}       
\bibliography{mypapers,ref}   

\begin{thebibliography}{10}
\providecommand{\url}[1]{{#1}}
\providecommand{\urlprefix}{URL }
\expandafter\ifx\csname urlstyle\endcsname\relax
  \providecommand{\doi}[1]{DOI \discretionary{}{}{}#1}\else
  \providecommand{\doi}{DOI \discretionary{}{}{}\begingroup
  \urlstyle{rm}\Url}\fi

\bibitem{Collins:1989gx}
J.C. Collins, D.E. Soper, G.F. Sterman, Adv. Ser. Direct. High Energy Phys.
  \textbf{5}, 1 (1989).
\newblock \doi{10.1142/9789814503266-0001}

\bibitem{Gribov:1972ri}
V.N. Gribov, L.N. Lipatov, Sov. J. Nucl. Phys. \textbf{15}, 438 (1972).
\newblock [Yad. Fiz.15,781(1972)]

\bibitem{Gribov:1972rt}
V.N. Gribov, L.N. Lipatov, Sov. J. Nucl. Phys. \textbf{15}, 675 (1972).
\newblock [Yad. Fiz.15,1218(1972)]

\bibitem{Dokshitzer:1977sg}
Y.L. Dokshitzer, Sov. Phys. JETP \textbf{46}, 641 (1977).
\newblock [Zh. Eksp. Teor. Fiz.73,1216(1977)]

\bibitem{Altarelli:1977zs}
G.~Altarelli, G.~Parisi, Nucl. Phys. \textbf{B126}, 298 (1977).
\newblock \doi{10.1016/0550-3213(77)90384-4}

\bibitem{Forte:1998nw}
S.~Forte, L.~Magnea, Phys. Lett. \textbf{B448}, 295 (1999).
\newblock \doi{10.1016/S0370-2693(99)00065-9}

\bibitem{Forte:2000wh}
S.~Forte, L.~Magnea, A.~Piccione, G.~Ridolfi, Nucl. Phys. \textbf{B594}, 46
  (2001).
\newblock \doi{10.1016/S0550-3213(00)00670-2}

\bibitem{Piccione:2001vf}
A.~Piccione, Phys. Lett. \textbf{B518}, 207 (2001).
\newblock \doi{10.1016/S0370-2693(01)01059-0}

\bibitem{Forte:2002us}
S.~Forte, J.I. Latorre, L.~Magnea, A.~Piccione, Nucl. Phys. \textbf{B643}, 477
  (2002).
\newblock \doi{10.1016/S0550-3213(02)00688-0}

\bibitem{Kotlorz:2006dj}
D.~Kotlorz, A.~Kotlorz, Phys. Lett. \textbf{B644}, 284 (2007).
\newblock \doi{10.1016/j.physletb.2006.11.054}

\bibitem{Kotlorz:2011pk}
D.~Kotlorz, A.~Kotlorz, Acta Phys. Polon. \textbf{B42}, 1231 (2011).
\newblock \doi{10.5506/APhysPolB.42.1231}

\bibitem{Kotlorz:2014kfa}
D.~Kotlorz, A.~Kotlorz, Phys. Part. Nucl. Lett. \textbf{11}, 357 (2014).
\newblock \doi{10.1134/S1547477114040153}

\bibitem{Kotlorz:2014fia}
D.~Kotlorz, S.V. Mikhailov, JHEP \textbf{06}, 065 (2014).
\newblock \doi{10.1007/JHEP06(2014)065}

\bibitem{Strozik-Kotlorz:2015gka}
D.~Strozik-Kotlorz, S.V. Mikhailov, O.V. Teryaev, PoS
  \textbf{BaldinISHEPPXXII}, 033 (2015)

\bibitem{Strozik-Kotlorz:2015iqr}
D.~Strozik-Kotlorz, S.V. Mikhailov, O.V. Teryaev, J. Phys. Conf. Ser.
  \textbf{678}(1), 012017 (2016).
\newblock \doi{10.1088/1742-6596/678/1/012017}

\bibitem{Psaker:2008ju}
A.~Psaker, W.~Melnitchouk, M.E. Christy, C.~Keppel, Phys. Rev. \textbf{C78},
  025206 (2008).
\newblock \doi{10.1103/PhysRevC.78.025206}

\bibitem{Wandzura:1977qf}
S.~Wandzura, F.~Wilczek, Phys. Lett. \textbf{B72}, 195 (1977).
\newblock \doi{10.1016/0370-2693(77)90700-6}

\bibitem{Burkhardt:1970ti}
H.~Burkhardt, W.N. Cottingham, Annals Phys. \textbf{56}, 453 (1970).
\newblock \doi{10.1016/0003-4916(70)90025-4}

\bibitem{Airapetian:2006vy}
A.~Airapetian, et~al., Phys. Rev. \textbf{D75}, 012007 (2007).
\newblock \doi{10.1103/PhysRevD.75.012007}

\bibitem{Adolph:2015saz}
C.~Adolph, et~al., Phys. Lett. \textbf{B753}, 18 (2016).
\newblock \doi{10.1016/j.physletb.2015.11.064}

\bibitem{Bjorken:1966jh}
J.D. Bjorken, Phys. Rev. \textbf{148}, 1467 (1966).
\newblock \doi{10.1103/PhysRev.148.1467}

\bibitem{Bjorken:1969mm}
J.D. Bjorken, Phys. Rev. \textbf{D1}, 1376 (1970).
\newblock \doi{10.1103/PhysRevD.1.1376}

\bibitem{Kumano:2004dw}
S.~Kumano, T.H. Nagai, J. Comput. Phys. \textbf{201}, 651 (2004).
\newblock \doi{10.1016/j.jcp.2004.05.021}

\bibitem{El-gendi:1969}
S.E. El-gendi, Comput. J. \textbf{12}, 282 (1969).
\newblock \doi{10.1093/comjnl/12.3.282}

\bibitem{Salam:2008qg}
G.P. Salam, J.~Rojo, Comput. Phys. Commun. \textbf{180}, 120 (2009).
\newblock \doi{10.1016/j.cpc.2008.08.010}

\bibitem{Kwiecinski:1995rm}
J.~Kwiecinski, Acta Phys. Polon. \textbf{B27}, 893 (1996)

\bibitem{Bartels:1995iu}
J.~Bartels, B.I. Ermolaev, M.G. Ryskin, Z. Phys. \textbf{C70}, 273 (1996)

\end{thebibliography}

%
%

\end{document}